\documentclass[aps,prb,twocolumn,showpacs,letterpaper]{revtex4}
\usepackage{graphicx}
\usepackage{subfigure}
\usepackage{amsmath}
\newcommand{\bfa}{{BaFe$_{2}$As$_{2}$}}

\newcommand{\bcfa}{{Ba(Fe$_{1-x}$Co$_x$)$_2$As}}

\begin{document}

\title{ Theoretical Investigation of Optical Conductivity in \bcfa } 


\author{A. Sanna$^1$, F. Bernardini$^2$, G. Profeta$^3$, S. Sharma$^1$, J. K. Dewhurst$^1$, A. Lucarelli$^4$, L. Degiorgi$^4$  , E. K. U. Gross$^1$ and  S. Massidda$^2$}

\affiliation{$^1$ Max-Planck Institute of Microstructure Physics, Weinberg 2, D-06120 Halle, Germany}
\affiliation{$^2$ CNR-IOM SLACS, and Dipartimento di Fisica, Universit\'a degli Studi di Cagliari, I-09042 Monserrato, Italy}
\affiliation{$^3$ CNISM-DIpartimento di Fisica, Universit\'a degli studi de L'Aquila, Via Vetoio 10, I-67010 Coppito, Italy}
\affiliation{$^4$Laboratorium f\"ur Festk\"orperphysik, ETH - Z\"urich, CH-8093 Z\"urich, Switzerland}

\begin{abstract}
We report on theoretical calculations of the optical conductivity of \bcfa, as obtained from density functional theory within the full potential LAPW method. 
A thorough comparison with experiment shows that we  are able to reproduce most of the observed experimental features, in particular a magnetic peak located at about 0.2 eV which we ascribe to antiferromagnetic ordered magnetic stripes. We also predict a large in-plane anisotropy of this feature, which agrees very well with measurements on detwinned crystals. The effect of Co doping as well as the dependence of plasma frequency on the magnetic order is also investigated.

\end{abstract}

\pacs{   }

\maketitle

The recent discovery of superconductivity in iron-based pnictide compounds \cite{Kamihara} has triggered intense experimental and theoretical research activity. 
A great deal of attention has been devoted to the relationship between superconductivity and magnetism. Similar to cuprates, the parent compounds exhibit antiferromagnetic spin-density-wave (SDW) order that disappears upon doping or pressure, giving rise to superconductivity.  
A possible coexistence of magnetism and superconductivity has also been observed\cite{Lumsden}, although this issue is still under debate\cite{Ni,Ning,Chu,Lumsden,Yin,Julien,Nandi,Gofryk,Marsik,Sangeeta,Monni}.

Despite the many similarities to cuprates,  important differences characterize the pnictides; most relevant being that their parent magnetic phase is not Mott insulating but rather metallic in nature. 
The preponderance of magnetic properties of FeAs compounds can be described by local/semilocal approximations to exchange correlation functionals within density functional theory (DFT),  
however the debate on the nature of magnetism (itinerant or localized) in this class of materials is still an open question\cite{MazinJohannes}. 
The degree and type of correlation in these materials is also under debate, due mostly to the ARPES data which suggests a strong renormalization of the Kohn-Sham energy bands\cite{Lu_ARPES}. 

The pairing mechanism in pnictides is not yet universally agreed upon -- but the possibility of standard phononic superconductivity has now been ruled out\cite{Boeri,Boeri2}. It has been suggested that the superconductivity in these materials could be mediated by magnetic excitations coupling electron and hole pockets of the Fermi surface and favouring $s$-wave order parameters, with opposite sign on different sheets of the Fermi surface ($s_{\pm}$ coupling).

What is clear is that the main physics in FeAs materials is controlled by a subtle interplay of magnetism and Fermi surface topology. One of the tools that can be used to probe these properties is optical measurement. 
In this work we report on our results of the optical properties of pure and Co-doped \bfa\ using DFT. A detailed comparison of this work with infrared optical conductivity experiments is also performed, and provides interpretation of most of the main features of the experimentally observed spectra\cite{Lucarelli,Lucarelli2,Chen,Chen2,Wu,Nakajima}.

\section {METHODOLOGY}
\label{sec.methods}

Our calculations were performed using the highly accurate all-electron full potential linearized augmented planewave method (LAPW)\cite{Singh_book} as implemented within the Elk\cite{elk} code.

All calculations presented here have been performed using the experimental crystal structure for undoped \bfa\cite{Rotter}, and not the low temperature orthorhombic distortion. The main spectral features for the doped \bfa\ have been checked to be independent of this choice. 

The optical conductivity tensor $\sigma(\omega)$ is related to the dielectric function
by\cite{GrossoPastori} (atomic units are used throughout):
\begin{align}
{\bf \epsilon}_{\alpha\beta}(\omega) =
 \delta_{\alpha\beta} + \frac{4\pi i}{\omega}\sigma_{\alpha\beta}(\omega),
\end{align}
for Cartesian components $\alpha,\beta$.
In practice, $\sigma$ is calculated according to\cite{RathgenKatsnelson} 
\begin{align}\label{eq:1}
 \sigma_{\alpha\beta}(\omega)&=-\frac{i}{\Omega}\sum_{vc{\bf k}}
 \frac{w_{\bf k}}{\varepsilon_{vc}}
 \left[\frac{\Pi_{vc}^{\alpha}({\bf k})\Pi_{cv}^{\beta}({\bf k})}{\omega+\varepsilon_{vc}+i\eta}
 +\frac{(\Pi_{vc}^{\alpha}({\bf k})\Pi_{cv}^{\beta}({\bf k}))^*}
 {\omega-\varepsilon_{vc}+i\eta}\right]\nonumber\\
 &+\frac{i(\omega_p)_{\alpha\beta}}{4\pi}\frac{1}{\omega+i\eta},
\end{align}
where the first and second terms are the interband and the intraband contributions respectively;
$\varepsilon_{vc}\equiv\varepsilon_v-\varepsilon_c$ are the differences between valence and conduction
eigenvalues; $w_{\bf k}$ are $k$-point weights over the Brillouin zone; $\Omega$ is the unit
cell volume; and
\begin{align}
 \Pi_{vc}({\bf k})\equiv\langle v{\bf k}|\hat{p}|c{\bf k}\rangle
\end{align}
are the momentum matrix elements.
The imaginary term $i\eta$ mimics a broadening of the spectrum owing to a finite lifetime,
and a value of 0.1 eV is used for this. The plasma frequency, which is also a tensor quantity,
is given by
\begin{align}
 (\omega_p^2)_{\alpha\beta}=\frac{4\pi}{\Omega}\sum_{n{\bf k}}w_{\bf k}\Pi_{nn}^{\alpha}({\bf k})
 \Pi_{nn}^{\beta}({\bf k})\delta(\varepsilon_n-\varepsilon_F),
\end{align}
with $\varepsilon_F$ being the Fermi energy.
We find the dielectric function and plasma frequency to be sufficiently well converged with a
mesh of $16\times16\times4$ $k$-points. 

\section {RESULTS AND DISCUSSION}
\label{sec.results}

\subsection{Undoped compound}
\subsubsection{Ground state properties}
We investigated the non-magnetic (NM) phase and various possible magnetic orderings. The most relevant is the striped phase (Str.), where on each Fe plane stripes of up spins are alternated with stripes of down spins.
We also studied the ferromagnetic (FM), checkerboard (CB) and non-collinear antiferromagnetic (2K) orderings. 
In the latter case, spins are rotated 90 degrees between adjacent Fe sites.
The corresponding total energies, relative to the Str. phase, have been collected in Table \ref{tab.data}.
In agreement with experiments and other DFT calculations \cite{JohannesMazinPRB}, we find the striped phase to be the most stable. 

\begin{table}[h]
\begin{center}
\begin{tabular}{l|ccccc}
                     &  NM    &  CB   &  Str.  & FM    & 2K    \\ \hline\hline
$E$ (meV)              &  120.0 &  70.1 &    0  & 119.5 & 33.1  \\
$M$(Fe) ($\mu_B)$      &    0   &  1.428& 1.490 & 0.31 & 1.33 \\ \hline
$\omega_p$ $x,y$      &  2.55  & 1.62  &{\footnotesize 0.70$x$/0.91$y$}   &{\footnotesize 1.958$\uparrow$/1.963$\downarrow$}& 1.44  \\
$\omega_p$ $z$ (eV)  &  0.79  & 1.54  & 1.03  &{\footnotesize 0.60$\uparrow$/0.87$\downarrow$}       & 2.349 \\ \hline
v$_F$ $x,y$           & 13.56  & 8.94  &{\footnotesize 7.77$x$/10.07$y$}   &{\footnotesize 12.61$\uparrow$/23.05$\downarrow$}       &  9.168   \\
v$_F$ $z$ ($10^4$m/s)&  4.22  & 8.52  & 11.45  &{\footnotesize 5.96$\uparrow$/10.26$\downarrow$}      &  14.915
      
\end{tabular}
\caption{Undoped \bfa\ data for different magnetic orderings. $E$ is the total energy per formula unit relative to the Str. phase; $M$(Fe) is the magnetic moment integrated in the Fe muffin tin; $\omega_P$ and v$_F$ are the plasma frequency and Fermi velocity respectively. In-plane components are labeled by $x,y$ and the out of plane component by $z$. For ferromagnetic ordering the symbol $\uparrow$ indicates the majority spin channel. In the Str. phase the spin stripes are oriented along the $x$ axis.}\label{tab.data}
\end{center}
\end{table}

As in the case of LaOFeAs\cite{Sangeeta}, the 2K structure is close in energy to that of the striped, while the FM solution is almost degenerate with the non-magnetic one. The magnetic moment, integrated in the Fe muffin tin, is larger than 1 $\mu_B$ in all three antiferromagnetic structures, and is equal to 1.49 $\mu_B$ in the ground state Str. phase (Tab. \ref{tab.data}). These theoretical moments are much larger than the experimental values measured by means of neutron diffraction ($\approx 0.9 \mu_B$\cite{Huang,Chen,Matan,Su}) and M\"ossbauer absorption(0.4 $\mu_B$\cite{RotterPRB}).

\subsubsection{Plasma frequencies}
 We first focus on the plasma frequency tensor ($\omega_{p}$), that determines the low frequency part of the spectrum.   
Owing to lattice symmetry, $\omega_{p}$ is diagonal with respect to the tetragonal axis, and we will indicate 
the $xx$, $yy$ and $zz$ components simply as $x$, $y$ and $z$.
 In all the investigated magnetic structures $\omega_{p}$ has only two independent components, apart from the Str. phase which has three. For the FM case we define separated plasma frequencies for majority and minority spin channels. The corresponding results are listed in Table \ref{tab.data}. 

The non magnetic phase yield a value $\omega_{p}$ of 2.55 eV. Magnetic order strongly affects the value of the plasma frequency in that magnetism opens a partial gap at $E_{F}$ thereby reducing both the Fermi velocity and the density of states at the Fermi energy\cite{SinghPRB} and resulting in a decrease of $\omega_p$ as compared to the non magnetic case.
The antiferromagnetic phases have the most isotropic $\omega_{p,z}/\omega_{p,x}$ ratio, while the NM structure is the most anisotropic. 
Our non-magnetic plasma frequencies are in agreement with previous calculations\cite{Qazilbash,Drechsler,Ferber}. Magnetic plasma frequencies in the Str. phase have been reported before in two recent works\cite{Tanatar,Ferber} and we find a reasonable agreement with our results. Slightly smaller and more isotropic values are reported in the past perhaps because of the larger value of the Fe moment (1.64 $\mu_B$ and 1.98 $\mu_B$).

The experimental in-plane plasma frequency, extracted from optical conductivity measurements, has been estimated in Ref. \onlinecite{Hu} to be $\omega_{p}=1.6$ eV at room temperature and 0.58 at 10 K. The low temperature limit is in good agreement with our estimate for the striped phase (our zero temperature ground state). 
On the other hand, the room temperature value,  measured above the N\'eel temperature ($T_N$), should be compared to our estimate in the non-magnetic phase (2.55 eV); the experimental value turns out to be much smaller. A temperature effect, not included in our calculations, could partially justify this disagreement. 
However a strong dependence of the plasma frequency on temperature is not expected: in case of a similar system, BaNi$_2$As$_2$, the plasma frequency drops by only 5\% of its value between 300 K and 150 K (above the structural/magnetic phase transition)\cite{ChenBaNi2As2}.

For the non-magnetic phase, a disagreement between LDA calculated plasma frequencies and experimental estimation from optical conductivity has been already reported, and been interpreted as an indication of correlation effects\cite{QazilbashArxiv,MazinArxiv,Drechsler,Qazilbash}.   
However, assuming that the main reason for the discrepancy is correlation, our results indicate that renormalization effects are much stronger in the \textit{normal} state than in the magnetic one. If we consider the ratio between the theoretical and experimental $\omega_p^2$, as a measure of the renormalization effect induced by correlations, as was done in Ref. \onlinecite{Qazilbash}, then quite unexpectedly we find the non-magnetic phase of \bfa\ to be a strongly correlated material, while magnetic \bfa\ appears to be a conventional one.

There has been disagreement in the literature on whether this implies that the dynamical spin correlation effects present in the metallic phase are frozen out in the magnetic one, or that the effect can not be ascribed to correlation effects at all, but is simply due to the inadequacy of the LSDA exchange-correlation functional (as suggested by Mazin in Ref. \onlinecite{MazinArxiv}). The veracity of either of these scenarios cannot be determined by the present calculations

\subsubsection{Optical Conductivity}

\begin{figure}[t]
\begin{center}
\includegraphics[clip=,width=0.49\textwidth]{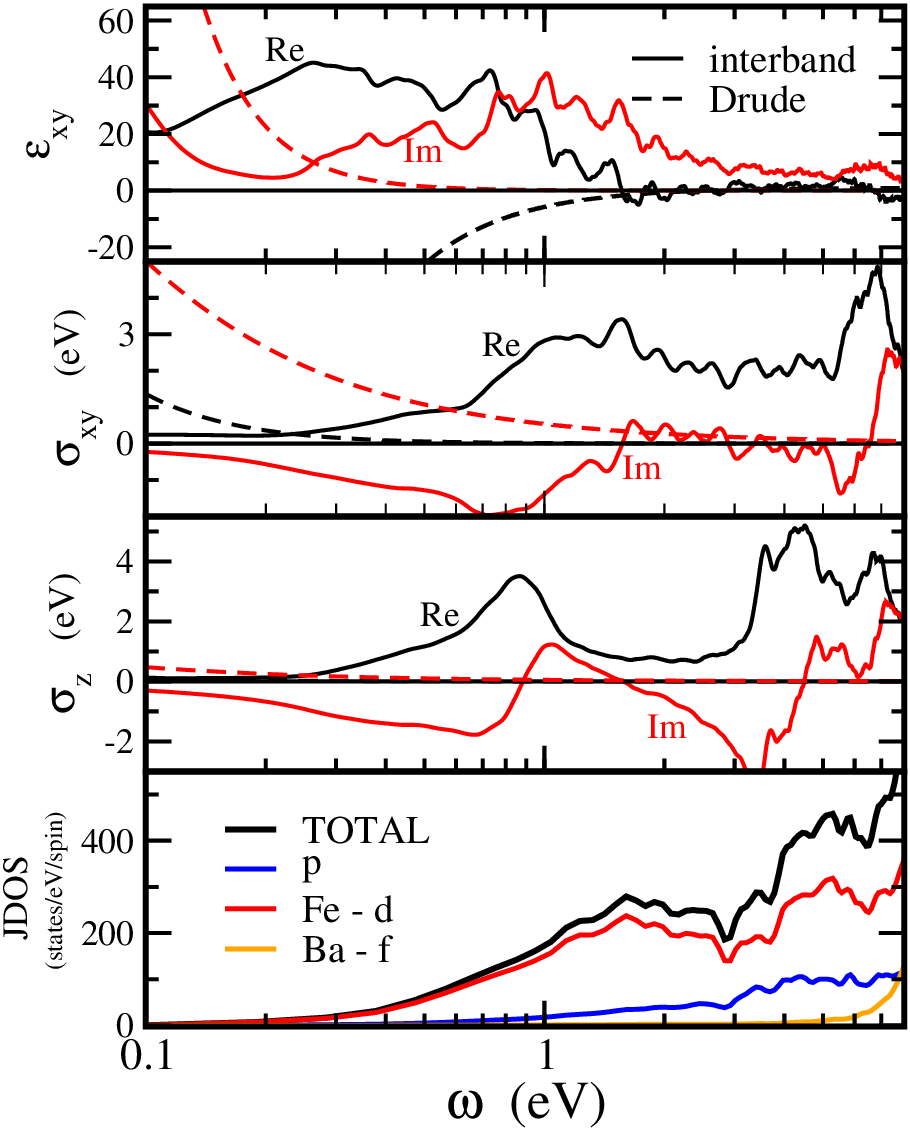}
\end{center}
\begin{minipage}{0.49\textwidth}
\begin{center}
\caption{(color online) Optical properties for non-magnetic \bfa. In the top panel Real (black) and Imaginary (red) components of the dielectric function. Dashed and full lines refer, respectively to intraband and interband contributions as in eq. \ref{eq:1}.
In the two central panels, the optical conductivity $\sigma_{x,y}$ refers to the in plane and $\sigma_z$ to the out-of-plane component. Dashed lines are the intraband contributions. The joint density of states is resolved into orbital contributions in the lowest panel.}\label{fig:sigmaP}
\end{center}
\end{minipage}
\end{figure}

Calculated optical conductivity for the non-magnetic phase is shown in the middle panels of Fig. \ref{fig:sigmaP}.  
Strong anisotropy between the in-plane (${\sigma_{xy}}$) and out of plane (${\sigma_z}$) conductivity is immediately apparent.  
In particular, the out-of-plane conductivity has a more pronounced frequency dependence than its in-plane counterpart. This anisotropy underscored by the appearance of a plasmonic peak, present only in $\sigma_z$. At this point it is worth noting that the value of the plasma frequency in the out-of-plane case is much smaller than in conventional metals, the reason for this lies in the low value of the Fermi velocity in the $z$ direction.

The origin of optical conductivity in the NM phase can be determined by looking at the joint density of states (JDOS), which is the real part of the optical conductivity with equal weight for all optical transitions (momentum matrix elements set to 1). This JDOS can be further resolved into various orbital contributions as follows:
\begin{equation}
{\rm JDOS}(E,l)=\sum_{n,n',{\bf k}}\delta\left(\varepsilon_{n{\bf k}}-\varepsilon_{n'{\bf k}}-E\right)w_{\bf k}P_{n'{\bf k}}(l)
\end{equation}
where $n$ are band indices and $P_{n{\bf k}}(l)$ is the projection of the Kohn-Sham state on the orbital channel $l$.
This orbital projected JDOS is shown in the lowest panel of Fig. \ref{fig:sigmaP}. Clearly, all the low energy features originate from the Fe $d$-states that dominate near the Fermi Energy.   

\begin{figure}[t]
\begin{center}
\includegraphics[clip=,width=0.49\textwidth]{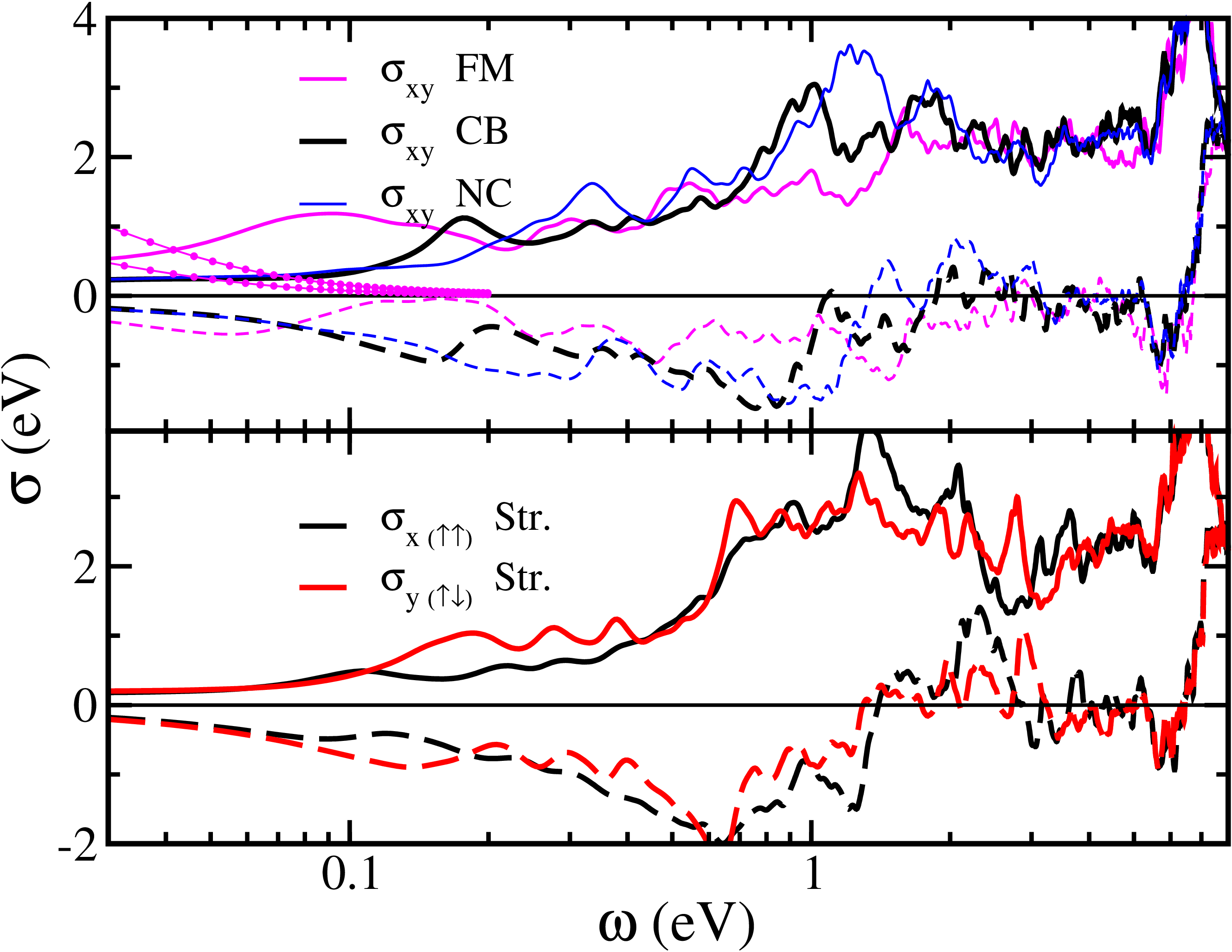}
\end{center}
\begin{minipage}{0.49\textwidth}
\begin{center}
\caption{(color online) In plane optical conductivity for various magnetic configurations of \bfa. Full and dashed lines give respectively real and imaginary parts of $\sigma$. All the data refer to the interband components only, apart from the FM case where the Drude contribution is also reported as a dotted line.
}\label{fig:sigma_M}
\end{center}
\end{minipage}
\end{figure}

The optical conductivity in various magnetic phases is shown in Fig. \ref{fig:sigma_M}. 
The main effect of magnetism is an increase of conductivity in the 0.1 - 1.0 eV range. The reason for this is that the reduced symmetry makes more states available (folding of the Brillouin zone due to the antiferromagnetic symmetry breaking), and new interband transitions are allowed in the low energy range.
As we will see in the following, the new structures merge in the Drude intraband contribution, as soon as the magnetic order disappears. 
This is anticipated by the behavior of $\sigma$ in the Ferromagnetic phase, where the ordered moment is low. In this phase a broad shoulder at about 0.1 eV is almost merged with the Drude peak (two dotted lines corresponding to the $\uparrow$ and $\downarrow$ channels). 
This enhancement of $\sigma (\omega)$  in the 0.1 - 0.3 eV region, induced by AF ordering, is an important feature of our results. This has been observed experimentally by different groups. We identify the peak at 1000 ${\rm cm}^{-1}$ (0.124 meV) reported in Refs. \onlinecite{Lucarelli,Lucarelli2,Wu,Nakajima} as a result of the AF ordering which introduces additional low frequency transitions. It is interesting to notice that in the striped phase it appears only for field polarization perpendicular to the stripes (let's call it \textit{off-stripe}). No similar effect exists when the electric field is in the direction of the stripes where the magnetic ordering is ferromagnetic. 

\begin{figure}[t]
\begin{center}
\includegraphics[clip=,width=0.49\textwidth]{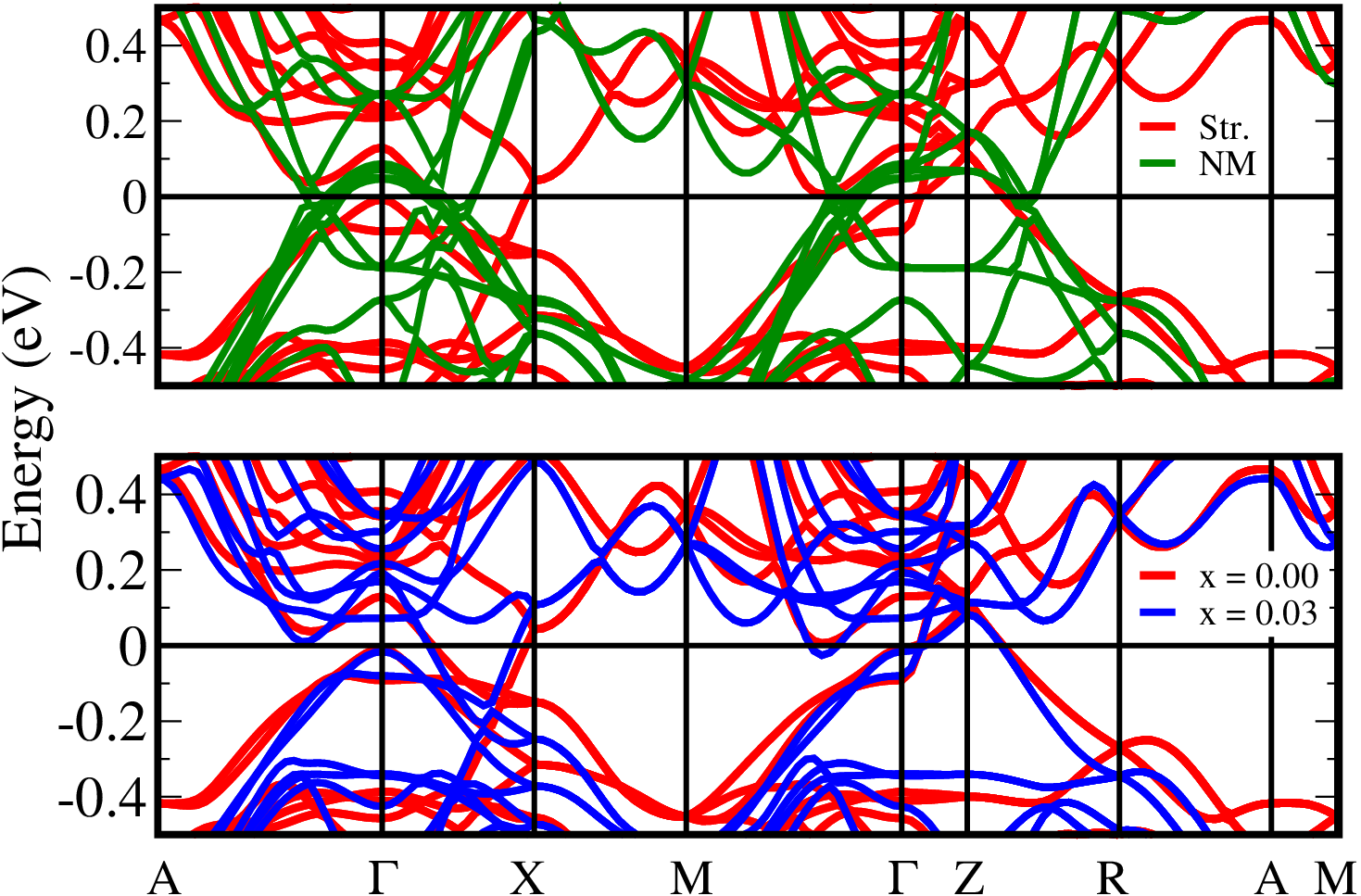}
\end{center}
\begin{minipage}{0.49\textwidth}
\begin{center}
\caption{(color online) Band structure near the Fermi energy (tetragonal unit cell). Upper panel:  comparison between non-magnetic and magnetic stripe ordered phase. Lower panel: in the striped phase, comparison between the undoped ($x=0.0$) and Co-doped ($x=0.03$) magnetic structure. }\label{fig:bands}
\end{center}
\end{minipage}
\end{figure}

In order to understand this we plot in Fig. \ref{fig:bands} the energy bands dispersion along the main lines of the Brillouin zone, including  the two orthogonal in-plane directions both in the top and basal planes. 
Compared to the usual representation of bands in this compound, we have here a folding of the Brillouin zone, with the $M$ point electron pockets folded back at $\Gamma$. 
We highlight in particular a Fe $d-$band located at about $-0.1$ eV at $\Gamma$, strongly dispersed in the direction of the stripe ($\Gamma-Y$) and flat in the off-stripe direction. The presence of this band offers a much larger phase-space availability for vertical transitions, compared to the non-magnetic
case and to the orthogonal direction. If this interpretation is correct, then the experimental peak can be fully understood in terms 
of single particle properties.

At higher energies, around 8 eV, a peak mostly due to Ba $f$-states appears in the imaginary part of $\sigma$, inducing a corresponding structure at about 6 eV in its real part. These high energy features are not affected by the magnetic ordering,  which induces important major changes only up to about 3 eV. 

\begin{figure}[t]
\begin{center}
\includegraphics[clip=,width=0.49\textwidth]{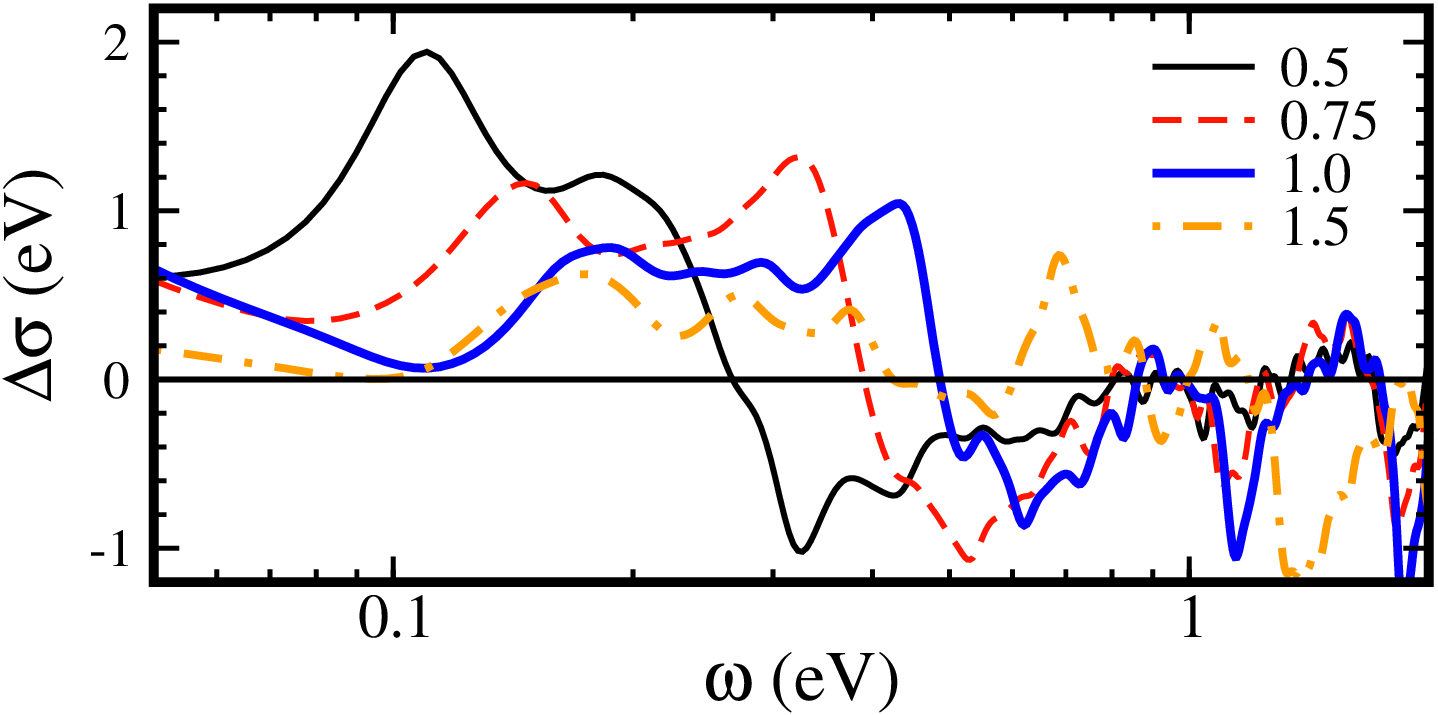}
\end{center}
\begin{minipage}{0.49\textwidth}
\begin{center}
\caption{(color online) In plane conductivity anisotropy in the striped phase of \bfa. Defined as the difference between conductivity in the off-stripe direction and stripe direction. Different curves correspond to a different magnetic moment at the Fe site (see legend for the moment value in atomic units). Drude contribution is included.}\label{fig:deltasigma}
\end{center}
\end{minipage}
\end{figure}

The effect of magnetic ordering can be singled out by defining the conductivity anisotropy $\Delta\sigma(\omega)=\sigma_y(\omega)-\sigma_x(\omega)$,  which is plotted in Fig. \ref{fig:deltasigma}. 
This anisotropy is strongly dependent on the magnetic moment, which in turn is strongly overestimated by standard DFT-LDA.  
In order to partially account for this effect, we perform fixed-spin-moment DFT calculations and report $\Delta\sigma$ as a function of the Fe magnetic moment varying from 1.5 $\mu_{B}$ (the value obtained in our ground state calculations) to zero. 

We see that for the calculated ground state, 
the low frequency part of the spectrum has an anisotropy peak centered around 0.25 eV ($\approx 2000 \  \rm cm^{-1}$). The spectral weight of this peak is compensated by depletion regions at high energy (see for example the negative peak around 1.5 eV). 
For smaller values of the magnetic moment this peak grows in intensity and also shifts toward lower frequencies. 
According to the above interpretation of the peak of $\sigma$, this can be understood in terms of the creation of a quasi-gap (depletion in the density of states) around the Fermi level $E_{F}$: a smaller moment implies a smaller quasi-gap, with a more peaked shape anisotropic structure.
The same happens to the high energy negative structures. 

Our theoretical spectrum and its anisotropy can be compared with the experimental results\cite{Lucarelli2}.
The presence of anisotropy peak, \emph{magnetic in origin}, is confirmed by experiment, both in its energy position and in its general shape, with a depletion of states around and above 500 meV (4000 $cm^{-1}$). 
To have a consistent comparison, we should consider our values calculated at the experimental magnetic moment. 
This is not easy as there is no universal agreement between different experiments: neutron diffraction and M\"ossbauer measurements give moments around 1 and 0.4 $\mu_{B}$ respectively\cite{Huang,Chen,Matan,Su,RotterPRB}. In light of this, our comparison  can only be semi-quantitative.  Still, using the results for $\mu_{Fe}=0.75-1 \mu_{B}$, the peak at 1000-2000  $cm^{-1}$ (120 - 250 meV) and the depletion of states after 3000-4000 $cm^{-1}$ (350 - 500 meV) found  in experiments are in fact well reproduced by theory.
We also notice that in experiments the peak position does not change with temperature, while in our case it does change with the magnitude of the  fixed moment. 
While the latter finding is clearly a consequence of the evolution of the band structure, the experimental behavior seems to point to a disorder of atomic moments, rather than to their reduced magnitude. 
\begin{figure}[t]
\begin{center}
\includegraphics[clip=,width=0.49\textwidth]{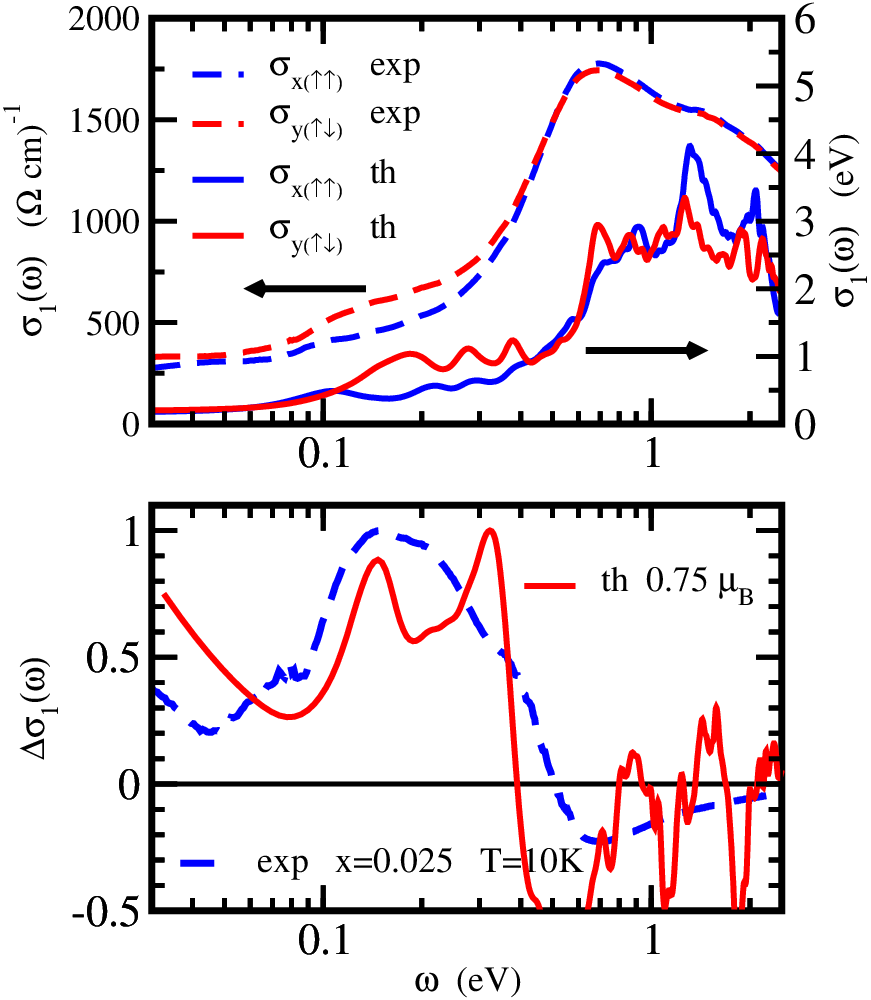}
\end{center}
\begin{minipage}{0.49\textwidth}
\begin{center}
\caption{(color online) Top panel, Experimental optical conductivity of detwinned \bcfa\ measured with the field along the two in plane Cartesian directions (red and blue dashed lines, compared with theoretical prediction in the magnetic Str. phase (full lines). 
Bottom panel, experimental optical conductivity anisotropy ($\sigma_y-\sigma_x$), at 0.025 of Co doping and 10 K, compared with the theoretical result obtained for a reduced magnetic moment of 0.75 $\mu_B$, close to experimental estimations (see text for details).
 }\label{fig:comparison}
\end{center}
\end{minipage}
\end{figure}

Bearing this in mind, we can make a detailed comparison between our results and experiments based on the anisotropy in the optical conductivity  recently observed on detwinned crystals\cite{Lucarelli2}. 
In Fig.~\ref{fig:comparison} we plot the experimental anisotropy together with theoretical results (Drude contributions are not included in both curves). We clearly see a fairly good agreement in the general shape. The main interband peak starts with a slightly broad shoulder at lower energy in experiments than the theoretical calculations. 
Most importantly, by comparing the $xx$ and $yy$ components of $\sigma$ we see 
very similar features in the magnetically induced anisotropy.  
The peak is found  at higher energy in our theoretical results; however, as seen in the previous discussion (see also Fig. \ref{fig:deltasigma}), this peak moves to lower energies as the Fe magnetic moment reduces. If we force the magnetic moment to be close to the experimental value (see lower panel in Fig.~\ref{fig:comparison}) the comparison improves 
significantly, with a very good agreement between theory and experiment.

\subsection{Co doping}

A fuller understanding of the optical properties of this system can be achieved by studying the doped system.
It is well established experimentally that electron or hole doping of the \bfa, rapidly destabilizes the magnetic phase.  
In the case of \bcfa\ at 6\% Co concentration magnetism disappears and the compound becomes a superconductor.
A possible coexistence between the two phases has been reported\cite{Christianson,Chu,Nandi}, possibly justified by nano-scale inhomogeneity in the dopants. In this work we assume that the magnetic phase is destroyed at 6\% of homogeneous Co doping, and that below this doping concentration superconductivity is absent. 

Keeping the crystal structure fixed, and using the virtual crystal approximation (VCA), this phase transition is not properly reproduced by DFT LSDA calculations\cite{SinghPRB}. The Fe moment seems to be only weakly reduced by electron doping (by about 20\% for x=0.06). 
The strong sensitivity of magnetism to the As  positions and the unavoidable impossibility of describing disorder effects could be all valid reasons for this disagreement with experiments.  This disagreement could again be due to correlation effects beyond LSDA.
We can simulate the experimental situation (and in particular the Co induced destabilization of magnetism) by doing fixed spin moment calculations, as before. At each doping (achieved using VCA) we fix the magnetic moment by scaling it according to the experimental magnetic critical temperature $T_N$\cite{Nandi,Chu}.
This constraint is likely to correct the deficiency of the conventional LSDA calculation and will mimic the two main Co induced effects: a localized electron doping on the Fe layer, and a magnetic to non-magnetic phase transition induced by this.

\begin{figure}[t]
\begin{center}
\includegraphics[clip=,width=0.49\textwidth]{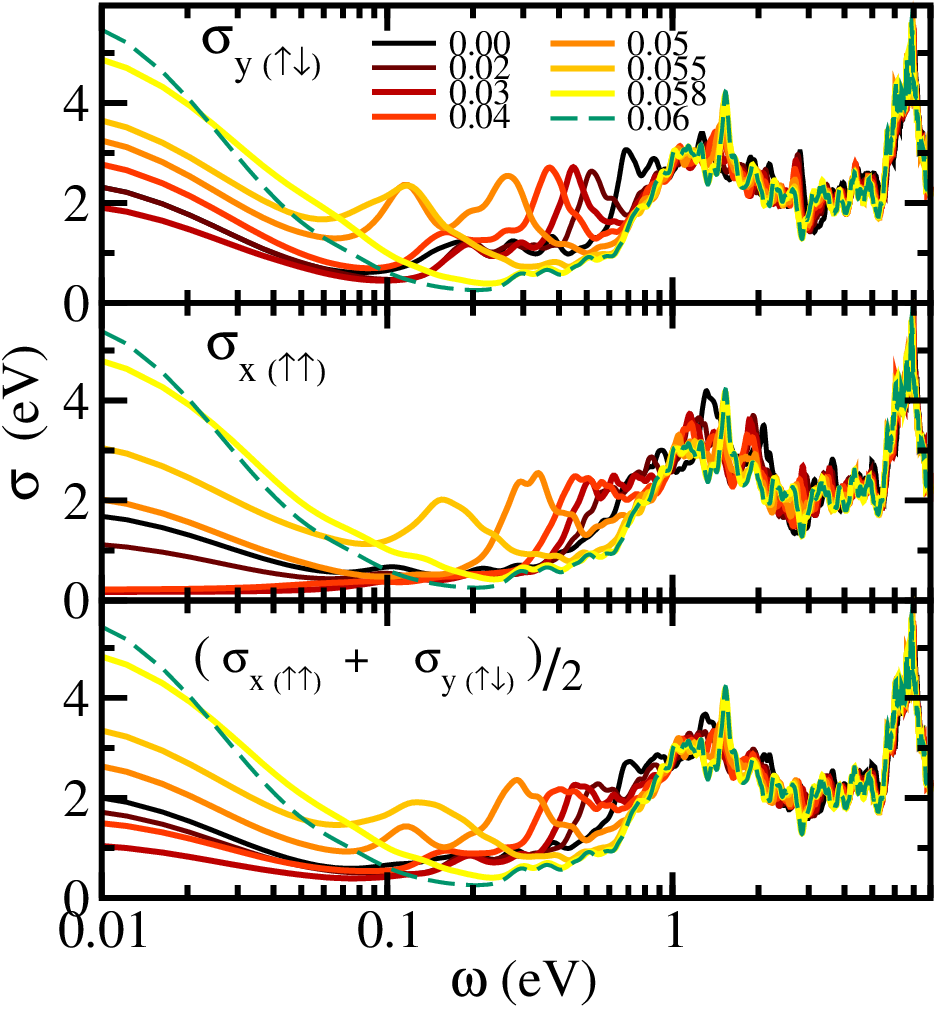}
\end{center}
\begin{minipage}{0.49\textwidth}
\begin{center}
\caption{(color online) In plane optical conductivity \bcfa, in the Str. phase.
 The x direction corresponding to the direction of the stripes. The x and y directions are averaged in the bottom panel, as they should come in an experimental measurement for twinned samples. Drude contribution is included.}\label{fig:Co_DOPING}
\end{center}
\end{minipage}
\end{figure}

The calculated optical conductivity in the Str. phase is plotted in Fig. \ref{fig:Co_DOPING}.
When doping reduces the Fe magnetic moment the system becomes more metallic --
an increase of the plasma frequency is accompanied by a large transfer of spectral weight from the structures around 0.8 eV to the Drude peak. The region of suppressed conductivity around 70 meV,  is filled as the magnetic gap closes. 
The observed trend is similar to the one reported in experiments (see for example the work of Nakajima \textit{et al.} \onlinecite{Nakajima} and Lucarelli \textit{et al.} \onlinecite{Lucarelli}). 
The main difference between our calculations and the experimental work (see for example Fig. [4] on Ref. \onlinecite{Nakajima}), is in the evolution of the spectra: we observe a transfer of spectral weight from one peak in the high energy part of the spectrum to the Drude component, while in experiments the spectral weight transfers  smoothly, without spectral peaks moving sharply as a function of $x$. 

We believe that this difference can be also understood in terms of the intrinsic inhomogeneity of the doping  in samples, present even in a macroscopically homogeneous Co distribution.
Experimental indications of this kind of inhomogeneity have been reported and discussed in several papers\cite{Ning,Gofryk}. Therefore, experimentally each spectrum would correspond, to some extent, to an average of a range of theoretical dopings, that are intrinsically homogeneous in the virtual crystal approximation.

\section{Conclusions}

In this work we report the calculated optical conductivity for \bfa, one of the most studied members of the iron arsenide family in order  to understand which experimental features can be reproduced by  theoretical methods based  on DFT. 
We confirm the fact that the plasma frequency in the non-magnetic phase is much higher than the experimentally observed values, indicating a poor description of the non-magnetic state within LSDA. On the other hand, in the magnetic striped phase the experimental estimate of the in plane plasma frequency (0.58 eV) is not far from our values (0.7 and 0.91 eV). If one ascribes the disagreement between theoretical and experimental results to correlation effects, as proposed by Qazilbash \textit{et al.}, this would mean that the virtual spin fluctuations freeze in at the onset of static magnetic ordering. 

Studying the frequency dependence of the optical conductivity, we observe that a low energy peak appears whenever the polarization of the incident light is directed along  an antiferromagnetically oriented spin configuration, as in the CB or Str. phase along the $y$ direction. This magnetic peak has been observed experimentally by various groups. In this work we are able to explain the experimentally observed large anisotropy of this peak\cite{Lucarelli2}.  

We also reported the effect of Co doping, by combining fixed moment calculations (to reproduce the experimental Fe moment) and the virtual crystal approximation and were able to reproduce the spectral weight transfer of the optical conductivity from the high to the low energy parts of the spectrum, as a function of such doping. This effect can be ascribed to the reconstruction of the electronic structure around the Fermi energy because of the magnetic ordering. 

\section{Acknowledgements}
F. B. acknowledges CASPUR for support under the Standard HPC Grant 2010 project. S. M. acknowledges support by the Italian MIUR through PRIN2008XWLWF9. 
A. L. and L. D. acknowledge support by the Swiss National Foundation for the Scientific Research within the NCCR MaNEP pool.


\end{document}